\documentclass[usenatbib]{mn2e}

\usepackage{amssymb,amsmath}
\usepackage{graphicx}
\usepackage{afterpage}
\usepackage{multirow}
\usepackage{natbib}
\usepackage{lineno}
\usepackage[usenames,dvipsnames]{color}
\def\citeN{\citet}
\def\cite{\citep}

\footnotesize
\newdimen\digitwidth    
\setbox0=\hbox{\rm0}
\digitwidth=\wd0
\catcode`!=\active
\def!{\kern\digitwidth}
\normalsize
\title[HTRU IV: Discovery \& polarisation of MSPs]{The High Time Resolution Universe Pulsar Survey IV: Discovery and polarimetry of millisecond pulsars}
\author[M.~J.~Keith et al.]
{M.~J.~Keith$^{1}$\thanks{Email: mkeith@pulsarastronomy.net},
S.~Johnston$^{1}$,
M.~Bailes$^{2,3,4}$,
S.~D.~Bates$^5$,
N.~D.~R.~Bhat$^{2,4}$,\newauthor
M.~Burgay$^6$,
S.~Burke-Spolaor$^{1}$,
N.~D'Amico$^{6}$,
A.~Jameson$^{2}$,
M.~Kramer$^{7,5}$, \newauthor
L.~Levin$^{2,1}$,
S.~Milia$^{6,8}$,
A.~Possenti$^6$,
B.~W.~Stappers$^5$,
W.~van Straten$^{2,4}$ and \newauthor
D.~Parent$^9$
\\
$^1$ Australia Telescope National Facility, CSIRO Astronomy \& Space Science, P.O. Box 76, Epping, NSW 1710, Australia\\
$^2$ Swinburne University of Technology, Centre for Astrophysics and Supercomputing Mail H39, PO Box 218, VIC 3122, Australia\\
$^3$ University of California, Berkeley, 601 Campbell Hall 3411, Berkeley, CA 94720, USA\\
$^4$ ARC Centre of Excellence for All-sky Astrophysics (CAASTRO)\\
$^5$ University of Manchester, Jodrell Bank Centre for Astrophysics, Alan Turing Building, Manchester M13 9PL, UK\\
$^6$ INAF-Osservatorio Astronomico di Cagliari, localit\`a Poggio dei Pini, strada 54, I-09012 Capoterra, Italy\\
$^7$ Max Planck Institut f\"ur Radioastronomie, Auf dem H\"ugel 69, 53121 Bonn, Germany\\
$^8$ Dipartimento di Fisica, Universit\`a degli Studi di Cagliari, Cittadella Universitaria, 09042 Monserrato (CA), Italy \\
$^9$ Center for Earth Observing and Space Research, College of Science, George Mason University, Fairfax, VA 22030, USA\\
}

\date{}
\begin{document}

\maketitle
\newcommand{\setthebls}{
}

\setthebls

\begin{abstract} 
We present the discovery of six millisecond pulsars (MSPs) in the High Time Resolution Universe (HTRU) survey for pulsars and fast transients carried out with the Parkes radio telescope.
All six are in binary systems with approximately circular orbits and are likely to have white dwarf companions.
PSR J1017--7156 has a high flux density and a narrow pulse width, making it ideal for precision timing experiments.
PSRs J1446--4701 and J1125--5825 are coincident with gamma-ray sources, and folding the high-energy photons with the radio timing ephemeris shows evidence of pulsed gamma-ray emission.
PSR J1502--6752 has a spin period of 26.7~ms, and its low period derivative implies that it is a recycled pulsar.
The orbital parameters indicate it has a very low mass function, and therefore a companion mass much lower than usually expected for such a mildly recycled pulsar.

In addition we present polarisation profiles for all 12 MSPs discovered in the HTRU survey to date.
Similar to previous observations of MSPs, we find that many have large widths and a wide range of linear and circular polarisation fractions.
Their polarisation profiles can be highly complex, and although the observed position angles often do not obey the rotating vector model, we present several examples of those that do.
We speculate that the emission heights of MSPs are a substantial fraction of the light cylinder radius in order to explain broad emission profiles, which then naturally leads to a large number of cases where emission from both poles is observed.

\end{abstract}


\newcommand{\polpsrsec}[1]{\subparagraph{#1}}
\section{Introduction}
Radio pulsars are typically classified in terms of the observed parameters of pulse period and its derivative, placing them on the `$P-\dot{P}$ diagram'.
They are generally thought to be born with spin periods of a few tens of milliseconds and magnetic field strengths of the order $10^{12}$\,G.
As the pulsar ages, the rotational energy is radiated away and the rotation rate slows, and therefore the pulsar traces a line across the $P-\dot{P}$ diagram, until it reaches the `death line' where radio emission ceases.
A distinct sub-group of radio pulsars, the ``millisecond pulsars'' (MSPs) typically have spin periods less than $\sim30$~ms, often just a few milliseconds.
These periods are shorter than the youngest pulsars, however their small period derivatives indicate spin-down ages of $>10^{9}$~years and they are most often found in binaries with white dwarf (WD) or neutron star (NS) companions.
The evolution of MSPs must therefore follow a different path to the majority of pulsars.
The accepted theory states that MSPs are initially formed in the same way as other pulsars, however after most are thought to have spun down to longer periods, they are then `recycled' via accretion of mass and angular momentum from a companion star~\cite{acrs82}.
This allows the pulsar to attain very short spin periods (although this is not always the case), and the accretion process is thought to also reduce the dipole magnetic field strength and therefore reduce the observed period derivative \cite{bk74,tv86,czb01}.

The majority of MSPs have spin periods less than 10~ms, and are either solitary or in circular orbits with a WD, or a very low mass companion ($< 0.03 M_\odot$).
Longer spin period MSPs tend to have larger mass companions.
Within  $10< P < 20$~ms the companions are always WDs, however above $20$~ms there are a broad range of systems including WD, NS \cite{bdp+03} and even Be star companions \cite{jml+92}.
At the time of writing, the pulsar catalogue\footnote{http://www.atnf.csiro.au/research/pulsar/psrcat/} lists 220 pulsars with a spin period\footnote{Even though recycling process can produce pulsars with spin periods longer than 30~ms, we choose this limit to easily distinguish MSPs from young pulsars.} less than 30~ms, of which only 95 are located outside of globular clusters.
Although the majority of the observed MSPs are located in globular clusters, the fact that many are formed via exchange interactions, and the effects of the cluster gravitational potential means that their evolutionary paths are different.
In this paper we only consider Galactic-field MSPs.
Only $2\%$ of pulsars with periods greater than 30~ms are in binaries, however there are 63 binary systems in the 95 non-cluster MSPs.

The pulse profiles of the majority of MSPs share many characteristics with those of the classically `young' pulsars, i.e. those with $P < 100$\,ms and $\dot{P} > 10^{-15}$~\cite{ts90}.
The profiles of MSPs are typically broad with polarisation fraction between $\sim 1\%$ and $\sim 50\%$ \cite{xkj+98,stc99,ovhb04,ymv+11}.
One way to understand the polarisation of pulsars is through the `rotating vector model' (RVM; \citealp{rc69a}) in which the observed polarisation angle is defined by the direction of a dipolar magnetic field at the point of emission.
The model is defined in terms of the geometric parameters $\alpha$, the angle between the rotation axis and the magnetic axis, and $\beta$, the angle between the magnetic axis and line of sight at closest approach.
For some pulsars, particularly those with components separated by $\sim 180^\circ$ this model has been used to measure these geometric parameters with high precision (e.g. \citealp{kj08,kjwk10}).
Although many MSPs have emission over a very wide phase window, sometimes greater than $180^\circ$, the position angles (PAs) often have large deviations from the best fit to the RVM, or the RVM simply does not model the observations well at all \cite{ovhb04,ymv+11}.

As well as the open questions intrinsic to the formation, evolution and emission of MSPs, they can also be valuable tools for other physics and astrophysics experiments.
Recently many radio MSPs have been discovered to have gamma-ray pulsations \cite{fermi_msps}, and indeed searches of {\it Fermi} Large Area Telescope (LAT) gamma-ray sources have led to the discovery of MSPs with both radio and gamma-ray emission \cite{rrc+11,cgj+11,kjr+11}.
This opens new possibilities for studying the broad-band emission phenomenon of pulsar and comparison between young pulsars and MSPs.
Although gamma-ray pulsations have been detected from some young pulsars without radio emission~\cite{fermi_16bsp,fermi_8bsp}, the gamma rays from most MSPs can only be binned into pulse profiles with the aid of a timing ephemeris obtained through radio timing.
The high precision time of arrival measurements that can be obtained from fast spinning MSPs, and their typically stable rotation rates mean that they can be used as precise clocks for tests of gravity and as broadband pulsed signals for probing the interstellar medium.
Of particular interest is the use of an array of MSPs to constrain the solar system ephemeris \cite{chm+10} and for the detection of low-frequency gravitational waves \cite{ych+11,vlj+11}.

Research into the origin of the  MSPs, the physics of their emission mechanism, and their use as tools for other astrophysical experiments are all facilitated by expanding the known sample.
To achieve this, the High Time Resolution Universe (HTRU) survey for pulsars and fast transients began at the Parkes radio telescope in 2008.
The high time and frequency resolution of the survey makes it much more sensitive to MSPs than previous efforts at Parkes \cite{kjv+10}.
Indeed, we have already announced the discovery of 6 MSPs \cite{bbb+11a,bbb+11b}.
In this paper we present the discovery and timing solutions for a further 6 MSPs, and discuss their origins (Section \ref{sec_disc}).
We also analyse {\it Fermi} LAT data associated with PSRs J1125--5825 and J1406--4701, both of which are coincident with LAT sources (Section \ref{sec_gamma}).
In Section \ref{sec_poln} we present high quality polarised profiles for all 12 MSPs discovered by the HTRU survey to date and discuss the implications in Section \ref{sec_discuss}.

\section{Discovery of six MSPs}
\label{sec_disc}
Here we report on the discovery of six MSPs in the HTRU survey, namely PSRs J1017--7156, J1337--6423, J1446--4701, J1502--6752, J1544--5149 and J1622--6617.
All six MSPs are in almost circular orbits, consistent with the standard model in which the spin-up of the pulsar is associated with Roche Lobe overflow from a nearby companion that circularised the orbit.
The companions are either white dwarfs or very low mass companions, such as those in orbit around PSR~J2051--0827 \cite{sbl+96}.
We note that all 12 MSPs discovered by the HTRU survey to date are part of binary systems, though there does not seem to be any obvious reason for the HTRU survey to preferentially select binaries.

\subsection{Observations and timing}
Each of these six pulsars was discovered in 540~s integrations as part of the mid-latitude portion of the HTRU survey being carried out at the Parkes radio telescope (see \citealp{kjv+10} for details).
Follow up observations were carried out at Parkes with the centre beam of the `HI-Multibeam' receiver and the Berkley-Parkes-Swinburne Recorder (BPSR), or once the orbital parameters were identified, the 3rd Parkes Digital Filterbank system (PDFB3).
The BPSR observations had an effective bandwidth of 340~MHz centred at 1352~MHz and the PDFB3 observations had a bandwidth of 256~MHz centred at 1369~MHz.
PDFB3 observations were calibrated for differential gain between the linear feeds by means of observation of a pulsed noise diode.
Each observation was averaged in time, frequency and converted to Stokes total intensity, then matched with an analytic reference profile to produce a single time-of-arrival measurement for each observation.
A model of the rotational, astrometric and orbital parameters was then fit to these measurements for each pulsar, in accordance with the standard pulsar timing procedure, using the {\sc tempo2} software \cite{hem06}.
Due to the almost circular nature of the orbits we have used the `ELL1' binary model as described in \citeN{lcw+01}. 
The best fit timing parameters and additional derived and observed properties of these six pulsars are presented in Tables \ref{discoveries1} and \ref{discoveries2}.
Flux densities presented are averaged over all 1369~MHz observations and are calibrated by comparison to the continuum radio source 3C\,218.
We do not quote a standard error in this value since all of these pulsars exhibit significant flux density variations due to interstellar scintillation.
Also shown in the table is the oft used gamma-ray detectability measure $\dot E^{1/2}/d^2$ \cite{fermi_pulsar_cat}.
MSPs detected by the {\it Fermi} satellite typically have $\dot E^{1/2}/d^2 > 10^{10}$\,erg$^{1/2}\,$pc$^{-2}\,{\rm s}^{-1/2}$ \cite{fermi_msps}.
The measured frequency derivative can be biased due to motion transverse to the line of sight \cite{shk70}.
No proper motion has yet been measured in these systems, however we can assume a typical velocity of 100\,km\,s$^{-1}$ \cite{tsb+99} and the DM derived distance to compute a likely fraction of the frequency derivative which could be due to this effect.
This is largest for PSR J1017--7156, with this value $\sim0.3\,\dot\nu$, however the other five pulsars have a likely Shklovskii contribution less than $0.2\,\dot\nu$.

\begin{table*}
\caption{
\label{discoveries1}
Observed parameters from timing of three previously unknown MSPs.
Values in parenthesis are the nominal 1-$\sigma$ {\sc tempo2} uncertainties in the last digits.
}
\begin{tabular}{llll}
\hline
Parameter & J1017$-$7156  & J1337$-$6423  & J1446$-$4701  \\
\hline
\hline
Right ascension, $\alpha$ (J2000) \dotfill & 10:17:51.32832(5)  & 13:37:31.928(18)  & 14:46:35.71432(13)  \\
Declination, $\delta$ (J2000) \dotfill & $-$71:56:41.64003(12)  & $-$64:23:04.88(3)  & $-$47:01:26.7616(15)  \\
$l$ ($^\circ$)\dotfill & 291.56 & 307.89 & 322.50 \\
$b$ ($^\circ$)\dotfill & $-12.55$ & $-1.96$ & 11.43 \\
Pulse frequency, $\nu$ (s$^{-1}$)\dotfill & 427.62190510627(9)  & 106.11873502(3)  & 455.64401644177(16)  \\
Frequency derivative, $\dot{\nu}$ (s$^{-2}$)\dotfill & $-$4.75(6)$\times 10^{-16}$  & $-$2.2(8)$\times 10^{-15}$  & $-$2.10(4)$\times 10^{-15}$  \\
Epoch of model (MJD)\dotfill & 55329.1  & 55234.7  & 55647.8  \\
Dispersion measure, DM (cm$^{-3}$pc)\dotfill & 94.2256  & 260.32  & 55.8340(11)  \\
\\
Binary model\dotfill & ELL1  & ELL1  & ELL1  \\
Orbital period, $P_b$ (d)\dotfill & 6.5118988121(19)  & 4.78533407(9)  & 0.2776660759(19)  \\
Projected semi-major axis, $a \sin i$ (lt-s)\dotfill & 4.83004527(11)  & 13.086499(11)  & 0.0640128(7)  \\
Epoch of ascending node, $T_{\rm asc}$ (MJD)\dotfill & 55329.10065316(5)  & 55234.770356(6)  & 55647.8044387(7)  \\
$e \cos \omega$\dotfill & $-$7.174(5)$\times 10^{-5}$  & 1.82(13)$\times 10^{-5}$  & 1.3(18)$\times 10^{-5}$  \\
$e \sin \omega$\dotfill & 1.2268(5)$\times 10^{-4}$  & 8.4(14)$\times 10^{-6}$  & 1.1(21)$\times 10^{-5}$  \\
Inferred eccentricity, $e$\dotfill & 1.4212(4)$\times 10^{-4}$ & $1.97(13) \times 10^{-5}$ & $<6\times 10^{-5}$ \\
Minimum companion mass, $m_{\rm c,min}$ (M$_\odot$)\dotfill & 0.19 & 0.78 & 0.019 \\
\\
Fit time span (MJD)\dotfill & 55343.2---55681.2  & 55460.0---55694.3  & 55358.6---55734.5  \\
RMS of residuals ($\mu s$)\dotfill & 0.8  & 37.5  & 2.5  \\
Reduced $\chi^2$\dotfill & 1.3  & 1.7  & 0.8  \\
\\
Mean flux density, $S_{1400}$ (mJy)\dotfill & 0.89 & 0.32 & 0.37 \\
Pulse width at $50\%$ of peak, $W_{50}$ ($^\circ$)\dotfill & 10 & 23 & 13 \\
Pulse width at $10\%$ of peak, $W_{10}$ ($^\circ$)\dotfill & 20 & 52 & 48 \\
Spin down energy loss rate, $\dot{E}$ (erg\,s$^{-1}$)\dotfill & $8.0\times 10^{33}$ & $9.2\times 10^{33}$ & $3.8\times 10^{34}$ \\
Characteristic age, $t_c$ (years)\dotfill & $1.4\times 10^{10}$ & $7.6\times 10^8$ & $3.4\times 10^9$ \\
Dipole magnetic field strength, $B_{surf}$ (G)\dotfill & $7.8\times 10^7$ & $1.4\times 10^9$ & $1.5\times 10^8$ \\
DM derived distance, $d$ (kpc)$^*$\dotfill & 3.0 & 5.1 & 1.5 \\
$\dot E^{1/2}/d^2$($\times 10^{10}$ erg$^{1/2}\,$pc$^{-2}\,{\rm s}^{-1/2}$)\dotfill & 1.0 & 0.5 & 9.0 \\
\hline
\end{tabular}
\end{table*}
\begin{table*}
\caption{
\label{discoveries2}
Observed parameters from timing of a further three previously unknown MSPs.
Values in parenthesis are the nominal 1-$\sigma$ {\sc tempo2} uncertainties in the last digits.
}
\begin{tabular}{llll}
\hline
Parameter & J1502$-$6752  & J1543$-$5149  & J1622$-$6617  \\
\hline
\hline
Right ascension, $\alpha$ (J2000) \dotfill & 15:02:18.610(4)  & 15:43:44.1498(10)  & 16:22:03.6669(6)  \\
Declination, $\delta$ (J2000) \dotfill & $-$67:52:16.78(2)  & $-$51:49:54.681(7)  & $-$66:17:16.978(4)  \\
$l$ ($^\circ$)\dotfill & 314.80 & 327.92 & 321.98 \\
$b$ ($^\circ$)\dotfill & $-8.07$ & 2.48 & $-11.56$ \\
Pulse frequency, $\nu$ (s$^{-1}$)\dotfill & 37.3909719910(4)  & 486.154232083(12)  & 42.33082901485(8)  \\
Frequency derivative, $\dot{\nu}$ (s$^{-2}$)\dotfill & $-$4.0(3)$\times 10^{-16}$  & $-$3.8(2)$\times 10^{-15}$  & $-$1.14(4)$\times 10^{-16}$  \\
Epoch of model (MJD)\dotfill & 55421.2  & 55522  & 55253.1  \\
Dispersion measure, DM (cm$^{-3}$pc)\dotfill & 151.75  & 50.92  & 87.94  \\
\\
Binary model\dotfill & ELL1  & ELL1  & ELL1  \\
Orbital period, $P_b$ (d)\dotfill & 2.4844570(5)  & 8.06077304(9)  & 1.640635183(20)  \\
Projected semi-major axis, $a \sin i$ (lt-s)\dotfill & 0.31756(3)  & 6.480281(5)  & 0.979380(6)  \\
Epoch of ascending node, $T_{\rm asc}$ (MJD)\dotfill & 55421.21202(4)  & 54929.067833(7)  & 55253.087284(3)  \\
$e \cos \omega$\dotfill & $-$5.3(125)$\times 10^{-5}$  & 2.02(11)$\times 10^{-5}$  & 4.3(117)$\times 10^{-6}$  \\
$e \sin \omega$\dotfill & $-$3.9(144)$\times 10^{-5}$  & 6.3(12)$\times 10^{-6}$  & 6.7(127)$\times 10^{-6}$  \\
Inferred eccentricity, $e$\dotfill & $< 2 \times 10^{-4}$ & 2.2(1)$\times 10^{-5}$ & $< 2\times 10^{-5}$ \\
Minimum companion mass, $m_{\rm c,min}$ (M$_\odot$)\dotfill & 0.022 & 0.22 & 0.092 \\
\\
Fit time span (MJD)\dotfill & 55360.4---55757.5  & 55432.3---55758.5  & 55256.9---55733.5  \\
RMS of residuals ($\mu s$)\dotfill & 67.9  & 13.5  & 22.2  \\
Reduced $\chi^2$\dotfill & 0.7  & 3.0  & 1.3  \\
\\
Mean flux density, $S_{1400}$ (mJy)\dotfill & 0.68 & 0.70 & 0.52 \\
Pulse width at $50\%$ of peak, $W_{50}$ ($^\circ$)\dotfill & 40 & 38 & 50 \\
Pulse width at $10\%$ of peak, $W_{10}$ ($^\circ$)\dotfill & -- & -- & 9.8 \\
Spin down energy loss rate, $\dot{E}$ (erg\,s$^{-1}$)\dotfill & $5.9\times 10^{32}$ & $6.1\times 10^{34}$ & $1.9\times 10^{32}$ \\
Characteristic age, $t_c$ (years)\dotfill & $1.5\times 10^9$ & $2.4\times 10^9$ & $5.9\times 10^9$ \\
Dipole magnetic field strength, $B_{surf}$ (G)\dotfill & $2.8\times 10^9$ & $1.7\times 10^8$ & $1.2\times 10^9$ \\
DM derived distance, $d$ (kpc)$^*$\dotfill & 4.2 & 2.4 & 2.2 \\
$\dot E^{1/2}/d^2$($\times 10^{10}$ erg$^{1/2}\,$pc$^{-2}\,{\rm s}^{-1/2}$)\dotfill & 0.2 & 4.2 & 0.3 \\
\hline
\end{tabular}
\end{table*}

\subsection{Two low mass binary pulsars: PSR~J1017--7156 and PSR~J1543--5149}
As seen in Tables \ref{discoveries1} and \ref{discoveries2}, PSR J1017--7156 and J1543--5149 have similar properties and fall into the category of `low mass binary pulsars' (LMBPs; \citealp{pk94,eb01}).
Other than the low mass function (companion mass $0.1 < M_{\rm c} < 0.5$ M$_\odot$), these pulsars are typified by an orbital period of a few days and a rapid rotation rate.
It should be noted however that there are a number of sources which fall into this category with longer spin periods \cite{eb01}.
LMBPs are by far the most common class of binary MSPs, with some 30 (of 57) such pulsars known.

PSR J1017--7156 has a spin period of 2.34~ms and a DM of 94.2~cm$^{-3}\,$pc and was discovered in an observation centred $\sim 0.07\degr$ from the nominal pulsar position, carried out on 2010-01-02 and detected with a signal-to-noise ratio (S/N) of 54.
The candidate was initially identified by a neural net system similar to that described in \citeN{emk+10}.
Follow up observations quickly identified the pulsar to be in a 6.5 day circular orbit around a low mass companion.
The properties of the binary are similar to those seen in other LMBPs.
The high S/N, short pulse period and narrow profile suggest that this pulsar is a good candidate for high precision timing.
Indeed, we can see that that this pulsar gives the best RMS residual of any HTRU discovery to date and it is already included as a target for the Parkes Pulsar Timing Array \cite{hbb+09}.

PSR J1017--7156 has a mean flux density of 0.89~mJy (the highest of any HTRU MSP), however even though its position was covered by previous surveys the relatively large DM ensured that it was not discernible from radio frequency interference due to excessive dispersion measure smearing.
To demonstrate this, folding the closest pointing in the \cite{ebvb01} survey shows a signal with a S/N of 11 and a pulse width of 1.3~ms which is broader than half the pulse period.
The superior frequency resolution of the BPSR back-end allowed for an easy detection with a pulse width of 0.16~ms width and a S/N of 54.
This clearly demonstrates the increased sensitivity of the BPSR back-end for pulsars with a high ${\rm DM}/P$ ratio.

PSR J1543--5149 has a pulse period of 2.06~ms, orbital period of 8 days and was first detected with a S/N of 16 in an observation carried out on 2010-08-24, offset $0.08\degr$ from the nominal pulsar position.
The minimum companion mass is 0.22 M$_\odot$, and so this pulsar is likely to be another member of the LMBP group.

\subsection{An intermediate mass binary pulsar: PSR~J1337--6423}
Intermediate mass binary pulsars (IMBPs) have heavier companion masses than the LMBPs and typically have spin periods of order 10~ms.
PSR J1337--6423 is a 9.4~ms pulsar with a DM of 260 cm$^{-3}\,$pc, the second largest DM of any MSP known to date.
It is located $0.04^\circ$ from the centre of the survey beam in which it was discovered with a S/N of 11.
The pulsar is in a 4.8~day orbit with an eccentricity of $<1\times 10^{-4}$.
These properties, and the minimum companion mass of 0.74~M$_\odot$ suggest that PSR J1337--6423 is likely to have a heavy WD companion and fall into the class of IMBPs.

\subsection{A very low mass binary pulsar: PSR~J1446--4701}
PSR J1446--4701 is a 2.2~ms pulsar in a compact binary with a 6.6~hour orbital period.
The pulsar was discovered in a survey beam centred $0.1^\circ$ from the nominal pulsar position on 2009-08-20 with a S/N of 12.4.
The minimum companion mass is just 0.019~M$_\odot$.
With the recent discovery of PSRs J0610--2100, J1731--1845, J2214+3000 and J2241--5236 \cite{bjd+06,bbb+11a,rrc+11,kjr+11}, all of which have very light companions, it is clear that the so-called ``Black Widow'' systems \cite{krst88,pebk88,vv88} are members of a larger class of very low mass binary pulsars (VLMBPs; \citealp{fck+03}).
These systems all show spin periods less than $\sim 4$~ms, orbital periods less than $\sim 10$ hours and a large $\dot E$.
Although many of these systems show eclipses for some of the orbit, as with PSR J0610--2100 and J2241-5236, PSR J1446--4701 does not show any sign of eclipse at 1369~MHz.

\subsection{Two intermediate-spin-period, low-mass binary pulsars: PSR~J1502--6752 and PSR~J1622--6617}
Although LMBPs typically have spin periods of a few ms, there are a handful of LMBPs that have spin periods greater than 10~ms.
Examples are PSRs J1745--0952 and J1841+0130, which have spin periods of 19 and 29~ms respectively \cite{eb01,lfl+06}, but are otherwise similar to the LMBPs.

PSR J1622--6617, with pulse period 23.6~ms, perhaps fits into the category of intermediate period LMBPs.
It was first observed on 2010-01-03, in a survey beam $0.07^\circ$ from the nominal pulsar position and with a S/N of 12.
We note that the observed spin and orbital parameters are similar to the known intermediate spin period LMBP J1841+0130 \cite{lfl+06}.

A more puzzling situation presents itself with the discovery of PSR J1502--6752.
Discovered with a S/N of 15.8 in an observation taken on 2010-06-11 (offset 0.08\degr\ from the nominal pulsar position) its spin period is 26.7~ms and the mass function implies a minimum companion mass of just 0.02~M$_\odot$, well below any known LMBP.
Indeed, PSR J1502--6752 and J1622--6617 have the smallest mass functions of any of the binary `intermediate period' MSPs (selected by $P>10$~ms and $\dot P < 10^{-18}$).
The true companion mass does, of course, depend on the inclination angle, and it is possible that we are observing these LMBPs close to face on.
For the companion mass of PSR J1502--6752 to be greater than 0.1~M$_\odot$, the inclination angle must be less than $13^\circ$, which has a probability of 0.025 of being drawn from a uniform distribution of three-dimensional orientations.
It is also worth noting that, as mentioned by \citeN{fck+03}, there is a distinct gap in the mass functions of the lowest mass binary MSPs between the VLMBPs, all of which have minimum masses less than 0.03~M$_\odot$ and the LMBPs of which all have masses greater than 0.09~M$_\odot$.
This suggests that there is a fundamental difference in the formation mechanism of the two systems.
The gap suggests that the LMBPs with the lowest mass functions must be somewhat inclined to the line of sight, and therefore we feel that the anomalously low mass function of PSR J1502--6752 is unlikely to be due to an inclination effect.

\begin{figure}
\includegraphics[width=8cm]{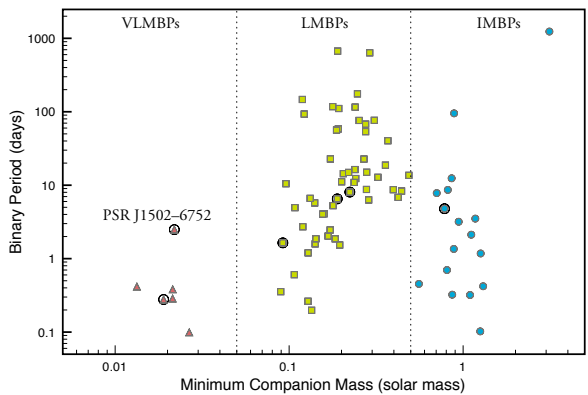}
\centering
\caption{
\label{m_pb}
Figure showing minimum companion mass and binary period for all known binary pulsars with spin period below 100~ms.
Pulsars with discoveries presented in this paper are circled.
We have split the binary MSP population into three categories as shown, the different symbols mark VLMBP (triangle), LMBP (square) and IMBPs (circle) as described in the text.
}
\end{figure}

To better understand the unique nature of J1502--6752, we can consider where it lies amongst the known population of MSPs.
By plotting the binary period and minimum companion mass of the known binary MSPs, as done in Figure \ref{m_pb}, we can identify at least three main populations of binary MSPs.
The VLMBPs appear in the lower left corner of the diagram, with masses lower than 0.03\,M$_\odot$, and binary periods which are typically shorter than a day.
The LMBPs contain the majority of the pulsars, and occupy a broad region in the centre of the plot, showing a weak trend between companion mass and binary period.
A third population, the IMBPs, have much more massive companions than LMBPs with similar binary periods.
There is some overlap between the LMBPs and IMBPs at binary periods greater than a few days, however it is clear that there is a distinct population for short period binaries.
We choose to split the LMBPs and IMBPs at by a minimum companion mass cut-off of 0.5\,M$_\odot$.
As described above, PSRs J1446--4701 and J1502--6752 fall into the VLMBP range, PSRs J1017--7156, J1543-5149 and J1622--6617 are LMBPs and PSR J1337-6423 is amongst those classified as IMBPs.

\begin{figure}
\includegraphics[width=8cm]{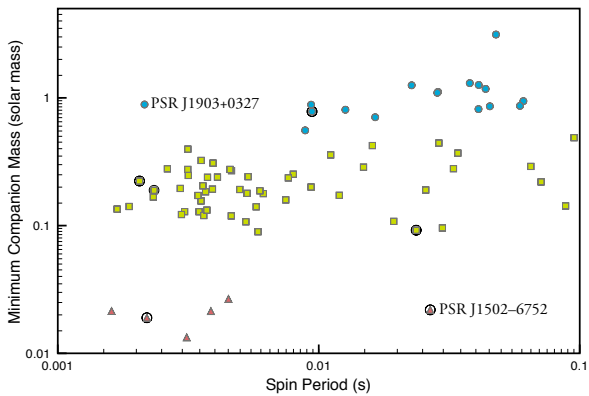}
\centering
\caption{
\label{p0_m}
Figure showing spin period and minimum companion mass for all known binary pulsars with spin period below 100~ms.
Pulsars with discoveries presented in this paper are circled.
The different symbols mark VLMBP (triangle), LMBP (square) and IMBPs (circle) as described in the text.
}
\end{figure}

Taking the same sample of pulsars, we now plot minimum companion mass against spin period, resulting in Figure \ref{p0_m}.
In this plane the IMBPs and LMBPs are much harder to distinguish, but we retain the classification from Figure \ref{m_pb}.
There are two clear outliers, the first is PSR J1903+0327, a known anomaly, which possesses a main sequence companion, and is probably the descendant of a triple system~\cite{crl+08,fbw+11}.
The second clear outlier is PSR J1502--6752, which sits alone in a large space in the lower right corner of the figure.
We must therefore consider if PSR J1502--6752 formed through the same channel as the VLMBPs, or if it underwent some unique process.

VLMBPs probably form when mass transfer begins
before a large degenerate core has formed as a result of the proximity
of the neutron star to the companion. This is a stable process,
and leads to a long period of spin-up and a low-mass companion
that is often prone to ablation and given the right geometry, can eclipse
the pulsar. How does PSR J1502$-$6752 fit into this scenario?
If we exclude the face-on orbit argument our only hint
comes from the orbital period of the system, which at 2.5\,d is
much greater than the other VLMBPs. Mass transfer in wider
orbits happens later in the donor star's life, leaving less time
for mass transfer. What is unclear is why similar orbital
period binaries have more evolved (and heavier) companions.
So while the spin period of PSR J1502$-$6752 is nicely
explained by the larger orbital period, the mass of the companion
is not. More speculative scenarios might involve the post-supernova
collision of the neutron star with a still main sequence companion
due to a favourably-oriented asymmetric kick \cite{bai89,tfv+99} and
the subsequent disruption of the host star and mild spin-up of
the pulsar.
The formation mechanism of VLMBPs is not yet well understood, and the discovery of PSR J1502--6752 undoubtedly complicates the picture further, although the uniqueness of this system may indeed suggest a process that is rare and improbable.
The discovery of further VLMBPs would be greatly beneficial in determining if PSR J1502$-$6752 really is unique, or if the spin period distribution of the VLMBPs is in fact wider than currently understood.

\section{Gamma-ray associations}
\label{sec_gamma}
Upon discovery of PSR J1446--4701 we determined that the gamma-ray source 2FGL J1446.8--4701 lies only 2.7\arcmin\ from the nominal radio position, well within the 95\% confidence radius of $r_{95}=7.2$\arcmin\ \cite{2FGL}.
From the radio timing, we measure the pulsar's spin-down energy loss rate $\dot E = 4\times 10^{34}$ erg\,s$^{-1}$, and $\dot E^{1/2} /d^2 = 9\times 10^{10}$~erg$^{1/2}\,$pc$^{-2}\,{\rm s}^{-1/2}$, assuming a DM-derived distance of 1.5~kpc.
This value is similar to that of other MSPs for which pulsed gamma rays have been detected \cite{fermi_pulsar_cat}.

Since the discovery of the HTRU MSP J1125--5825 \cite{bbb+11a}, the gamma-ray source 2FGL J1125.0--5821 has been discovered, lying only 6\arcmin\ away, also within the gamma-ray error radius, $r_{95}$=9\arcmin \cite{2FGL}.
The energy loss rate of the pulsar is $\dot E = 7.9\times 10^{34}$ and $\dot E^{1/2} /d^2 = 4 \times 10^{10}$~erg$^{1/2}\,$pc$^{-2}\,{\rm s}^{-1/2}$, assuming a DM-derived distance of 2.6~kpc \cite{bbb+11a}.

\begin{figure}
\includegraphics[width=8cm]{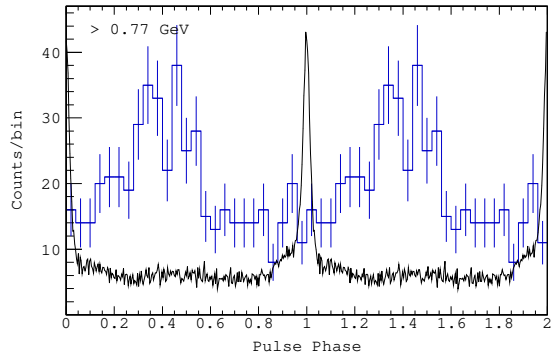}
\centering
\caption{
\label{1446_gamma}
Result of folding the LAT gamma-ray photons with the radio ephemeris of PSR J1446--4701 (histogram), overlaid with the 1369~MHz radio profile (curve).
The profiles are aligned assuming a DM of 55.83~cm$^{-3}$\,pc, and an arbitrary scaling for the radio flux density.
}
\end{figure}

\begin{figure}
\includegraphics[width=8cm]{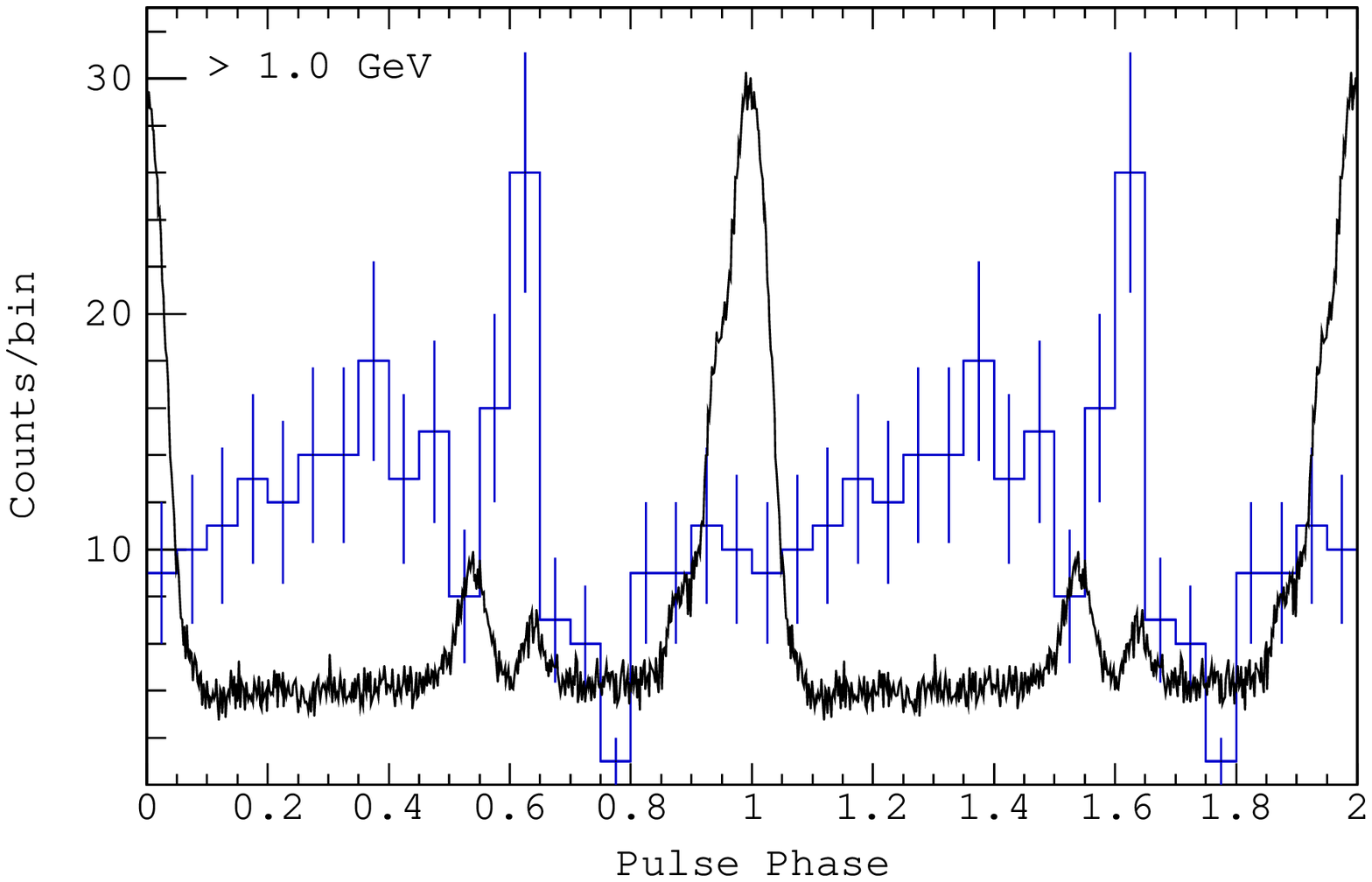}
\centering
\caption{
\label{1125_gamma}
Result of folding the LAT gamma-ray photons with the radio ephemeris of PSR J1125--5825 (histogram), overlaid with the 1369~MHz radio profile (curve).
The profiles are aligned assuming a DM of 124.79~cm$^{-3}$\,pc, and an arbitrary scaling for the radio flux density.
}
\end{figure}

We searched for gamma-ray pulsations from both of these pulsars by phase binning {\it Fermi} LAT photons (arriving between MJD 54683 and 55774) using the radio ephemeris.
We used ``Pass-6 diffuse" class events (highest probability of being gamma-ray photons) and excluded events with zenith angles $> 100^\circ$ to reject atmospheric gamma rays from the Earth's limb.
We used the bin-independent H-test statistic \cite{htest2010} to search over two parameters: the maximum angular separation from the pulsar position ($0.1^\circ < r < 2^\circ$), and the minimum photon energy cutoff ($100 < E_{\rm cutoff} < 1000$ MeV), with a maximum energy fixed at 50~GeV.
This truncates the point-spread function at low energies and decreases the number of background events.
The best signal for PSR J1446--4701 ($r = 0.8^\circ$, $E_{\rm cutoff} > 770$ MeV) is shown in Figure \ref{1446_gamma} with a post-trials significance of $5.4$-$\sigma$, while the best signal for PSR J1125--5825 ($r = 0.4^\circ$, $E_{\rm cutoff} > 1000$ MeV) is presented in Figure \ref{1125_gamma} with a post-trials significance of $4.9$-$\sigma$.

The alignment with the radio profile is accurate to about 0.01 in pulse phase as the DM is well constrained.
The gamma-ray profile of PSR J1446--4701 is fairly broad and has its peak at a phase of 0.5 relative to the narrow radio peak.
Although much lower in S/N, the radio and gamma-ray profile shape and alignment are somewhat similar to PSR J0437--4715 \cite{fermi_msps}.
In PSR J1125--5825 the gamma-ray peak trails the radio peak by a phase of 0.6, which is large, but not uncommon for millisecond pulsars with complex radio profiles.
The gamma-ray profile may be similar to PSR J2124--3358, which also shows a very complex radio profile, or perhaps similar to the double peaked profiles such as PSR J0030+0451 but with an absent leading peak \cite{fermi_msps}.

\section{Polarisation profiles}
\begin{table*}
\caption{
\label{rmtable}
Pulse period ($P$), DM and derived energy loss rate ($\dot{E}$) obtained from radio timing of the 12 MSPs discovered in the HTRU survey to date.
Note that $\dot{E}$ is not corrected for the Shklovskii effect, which may be large in some pulsars.
Also provided is the rotation measure (RM), pulse width at 50\% and 10\% of the peak flux density ($W_{50}$ and $W_{10}$), as derived from the 1369~MHz polarised pulse profiles described in this work.
References for discovery and full parameters are: [a] this work, [b] \citeN{bbb+11a} and [c] \citeN{bbb+11b}.
}
\begin{tabular}{lllllllll}
\hline
PSR & $P$ (ms) & DM (cm$^{-3}$pc)& $\dot{E}$ ($\times 10^{33}$ erg\,s$^{-1}$) & RM (rad\,m$^{-2}$) & W$_{50}$  ($^\circ$) & W$_{10}$  ($^\circ$) & Ref.\\
\hline
\hline
J1017--7156 & 2.34 & 94.2  & $8.0$ &            $-78(3)$   & 11 & 24 & [a]\\
J1125--5825 & 3.10 & 124.8 & $79$ &            $-7(3)$    & 32 & 200 & [b]\\
J1337--6423 & 9.42 & 260.3 & $9.2$ &            $-135(17)$ & 23 & 52 & [a]\\
J1446--4701 & 2.19 & 55.8  & $38$ &            $-14(3)$   & 13 & 48 & [a]\\
J1502--5143 & 26.7 & 151.8 & $0.59$ &            $-225(2)$  & 40 & 230 & [a]\\
J1543--5149 & 2.06 & 50.9  & $61$ & \phantom{$-$}0(25)    & 38 & 200 & [a]\\
J1622--6617 & 23.6 & 88.0  & $0.19$ & \phantom{$-$}70(30)   & 9.8 & 50 & [a] \\
J1708--3506 & 4.51 & 146.8 & $9.9$ &            $-15$(5)   & 60  & 139 & [b]\\
J1719--1438 & 5.79 & 36.8  & $1.5$ & \phantom{$-$}16(4)    & 20  & 63 &[c]\\
J1731--1845 & 2.34 & 106.6 & $76$ & \phantom{$-$}19(1)    & 11  & 28 &[b]\\
J1801--3212 & 7.45 & 176.7 & $0.25$ & \phantom{$-$}226(4)   & 25  & 120 &[b]\\
J1811--2404 & 2.66 & 60.6  & $28$ & \phantom{$-$}23(3)    & 14 & 280 & [b]\\
\hline
\end{tabular}
\end{table*}

\label{sec_poln}
We now consider the polarisation properties of all 12 MSPs discovered in the HTRU survey to date \cite{bbb+11a,bbb+11b}, the basic parameters of which are given in Table \ref{rmtable}.
To form high S/N profiles we summed all the observations used for timing of each pulsar with 64, 256 and 1024~MHz wide bands centred at 732, 1369 and 3100~MHz respectively.
The signals were passed through a poly-phase filterbank and folded on-line using the Parkes Digital Filterbank System (PDFB3 and PDFB4).
In the case of PSR J1017--7156 the ATNF-Parkes-Swinburne Recorder (APSR) was used to obtain coherently dedisperesd profiles at 732 and 1369~MHz.
Observations at 732 and 3100~MHz were performed with the ``10-50'' receiver.
Observations at 1369~MHz were performed with the centre beam of the ``HI-Multibeam'' receiver.
Before each observation a noise diode coupled to the receptors in the feed is observed to calibrate for differential gain and phase between the feeds.
To correct for cross-coupling of the receptors in the Multibeam receiver, we used a model of the Jones matrix for the receiver computed by observation of the bright pulsar PSR J0437--4715 over the entire range of hour angles visible, using the `measurement equation modelling' technique described in \citeN{van04c}.
The majority of timing is carried out at 1369~MHz, therefore profiles at the other frequencies are only included if the S/N ratio was high enough.
Note that we follow the `RVM sign convention' for the PA (see \citealp{ew01}).
We measure the Faraday rotation observed towards each pulsar by fitting PA variations across the 256~MHz band centred at 1369~MHz, using the algorithm described in \citeN{njkk08}.
The best fit values of rotation measure (RM) are provided in Table \ref{rmtable}.

We now present Figures \ref{J1017-7156}-\ref{J1811-2404}, showing profiles of the 12 MSPs in total intensity, linear and circular polarisation.
When multi-frequency data are present we include profiles at each of the frequencies, with an arbitrary phase alignment.
Additionally we show the polarisation PAs, corrected with the RM from Table \ref{rmtable}, and therefore absolute PAs as emitted at the pulsar, under the assumption that the RM used is correct.
The figures are marked with a horizontal bar indicating the amount of uncorrected dispersion smearing due to the finite channel width of the DFBs.

\subsection{Description of profiles}

\begin{figure}
\includegraphics[width=7cm]{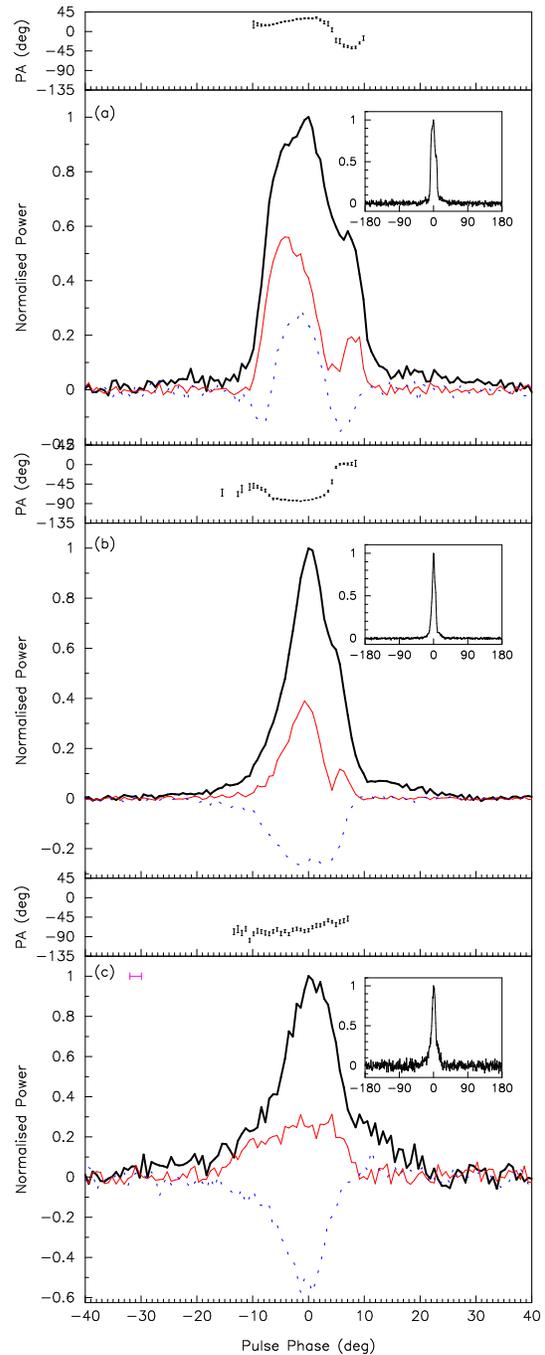}
\centering
\caption{
\label{J1017-7156}
Polarisation profiles of PSR J1017--7156 at (a) 732, (b) 1369 and (c) 3100~MHz, showing phases within $\pm 40^\circ$ of the pulse peak.
The solid black line shows total intensity, the thinner solid line shows linear polarisation and the dotted line shows circular polarisation.
The inset figures show the profile over the full $360^\circ$ of pulse phase.
The data at 732 and 1369~MHz have been coherently dedispersed and so do not exhibit any DM smearing.
}
\end{figure}

\polpsrsec{PSR J1017--7156} (Figure \ref{J1017-7156}).
The profile of PSR J1017--7156 exhibits a single emission region with width at half the peak value of $10^\circ$ at 1369~MHz, and a slight broadening at low frequencies.
In general, the frequency evolution of the profile is small, typical for MSPs \cite{kll+99}, although there is some evolution noticeable on the leading edge between 732 and 1369~MHz.
The linear polarisation fraction is about $30\%$ at all three frequencies, however the circular polarisation goes from $20\%$ at 732~MHz to $30\%$ at 1369~MHz to $45\%$ at 3100~MHz.
At 732~MHz the circular polarisation is predominantly positive, however by 1369~MHz it is entirely negative in sign.
Combined with the PAs (assuming a rotation measure of $-76$~rad\,m$^{-2}$), this strongly suggests that the emission at 732~MHz is dominated by a polarisation mode orthogonal to that which is dominant at 1369 and 3100~MHz.
The polarisation PAs also show considerable variation with frequency, especially towards the trailing edge of the profile.
At 732~MHz the PAs show a downwards `S-curve`, however at 1369~MHz the PAs exhibit a `U' shape, and at 3100~MHz, the PAs show an upwards slope.
We discuss the atypical properties of the polarised emission observed from this pulsar in Section \ref{1017_discuss}.

\begin{figure}
\includegraphics[width=7cm]{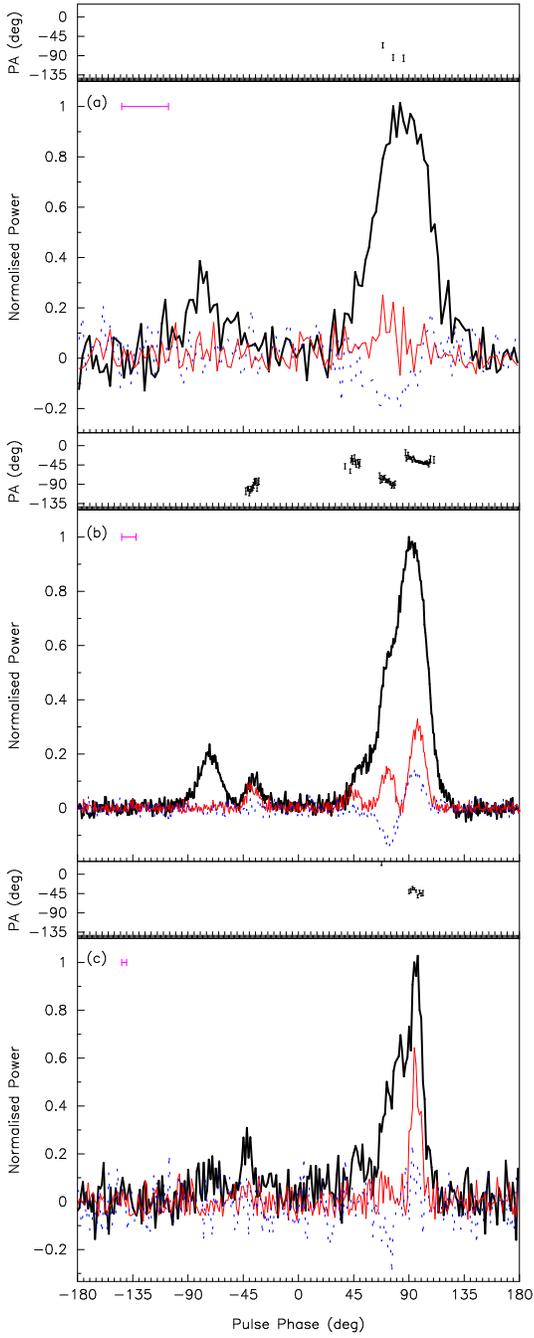}
\centering
\caption{
\label{J1125-5825}
Polarisation profiles of PSR J1125--5825 at (a) 732, (b) 1369 and (c) 3100~MHz.
}
\end{figure}

\polpsrsec{PSR J1125--5825} (Figure \ref{J1125-5825}).
At 1369~MHz, the profile of PSR J1125--5825 exhibits three distinct components.
Though the S/N is much lower, the profile shape is similar at other frequencies, with some dispersion broadening apparent at low frequencies.
At 1369~MHz the brightest of the three main components appears to be composed of at least three overlapping sub-components.
At the higher frequency it appears that this component is starting to break up into three or four narrower peaks..
In total, emission extends over $\sim 200^\circ$ of the profile.
The brightest component exhibits varying linear and circular polarisation, with an orthogonal mode jump aligned with the centre of the peak.
There is also a change in the handedness of circular polarisation associated with the jump.
The other two components show no circular polarisation, one is $100\%$ linearly polarised and the other is completely unpolarised.
The spectral index of the components at phase $-45^\circ$ is flatter than that of the component at $-80^\circ$.

\begin{figure}
\includegraphics[width=7cm]{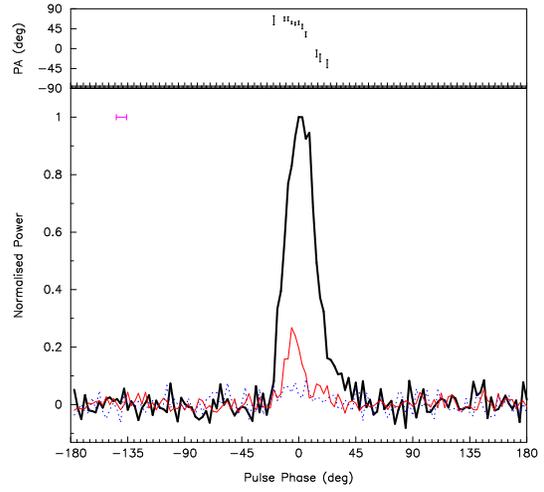}
\centering
\caption{
\label{J1337-6423}
Polarisation profile of PSR J1337--6423 at 1369~MHz.
}
\end{figure}

\polpsrsec{PSR J1337--6423} (Figure \ref{J1337-6423}).
The emission of PSR J1337--6423 extends over $\sim 90^\circ$ of pulse phase, with a half-width of $23^\circ$.
This profile is $\sim 5\%$ polarised in both linear and circular, with the polarised intensity roughly following the shape of the total intensity.
The PA shows a typical `S-shaped' swing.

\begin{figure}
\includegraphics[width=7cm]{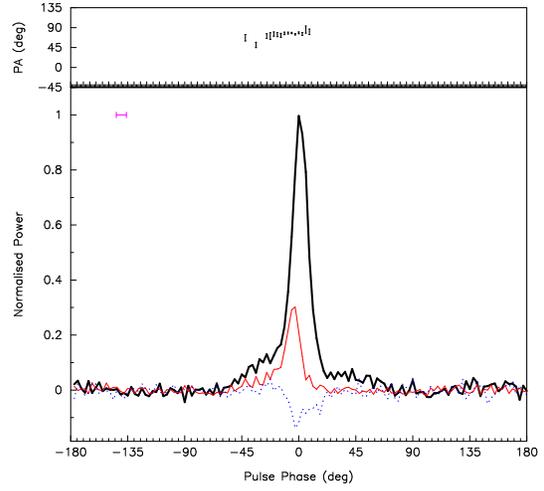}
\centering
\caption{
\label{J1446-4701}
Polarisation profile of PSR J1446--4701 at 1369~MHz.
}
\end{figure}

\polpsrsec{PSR J1446--4701} (Figure \ref{J1446-4701}).
The profile of PSR J1446--4701 is fairly simple, showing a single peak with half width $\sim 13^\circ$ lead by a small shoulder.
The profile has $\sim 20\%$ linear and circular polarisation, with the polarised intensity roughly following the shape of the total intensity.
The polarisation PAs are flat over the entire profile.
The on the trailing edge of the peak (phases 0-20$^\circ$), the ellipticity of the polarisation increases since the circular polarisation remains constant whilst the linear polarisation drops sharply.

\begin{figure}
\includegraphics[width=7cm]{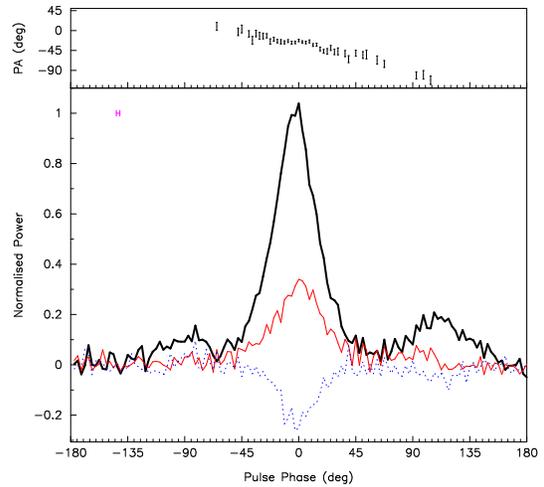}
\centering
\caption{
\label{J1502-6752}
Polarisation profile of PSR J1502--6752 at 1369~MHz.
}
\end{figure}

\polpsrsec{PSR J1502--6752} (Figure \ref{J1502-6752}).
The profile of PSR J1502--6752 covers nearly $300^\circ$ of pulse phase.
The main component has a polarisation fraction of $\sim 40\%$ linearly and $\sim 30\%$ circular.
The remainder of the profile does not have sufficient S/N to determine the polarisation fraction.
The observed PAs appear to change linearly with pulse phase.

\begin{figure}
\includegraphics[width=7cm]{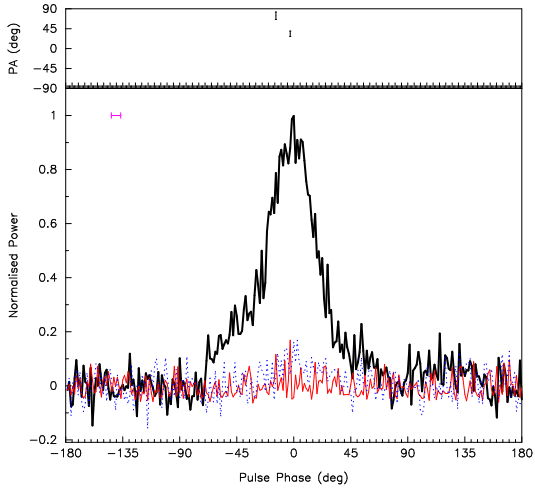}
\centering
\caption{
\label{J1543-5149}
Polarisation profile of PSR J1543--5149 at 1369~MHz.
}
\end{figure}

\polpsrsec{PSR J1543--5149} (Figure \ref{J1543-5149}).
The profile of PSR J1543--5149 is composed of a single, broad, peak with small shoulders on the leading and trailing edges.
There is less than $5\%$ polarisation fraction for both linear and circular.

\begin{figure}
\includegraphics[width=7cm]{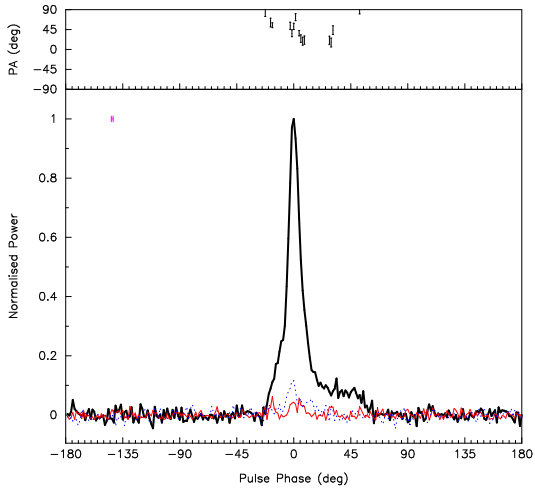}
\centering
\caption{
\label{J1622-6617}
Polarisation profile of PSR J1622--6617 at 1369~MHz.
}
\end{figure}

\polpsrsec{PSR J1622--6617} (Figure \ref{J1622-6617}).
The profile of PSR J1622--6617 is composed of a narrow peak with a broad shoulder on the trailing edge.
The polarisation fraction is small throughout, however the linear and circular polarisation track each other well, with a $\sim 10\%$ polarisation fraction in each.

\begin{figure}
\includegraphics[width=7cm]{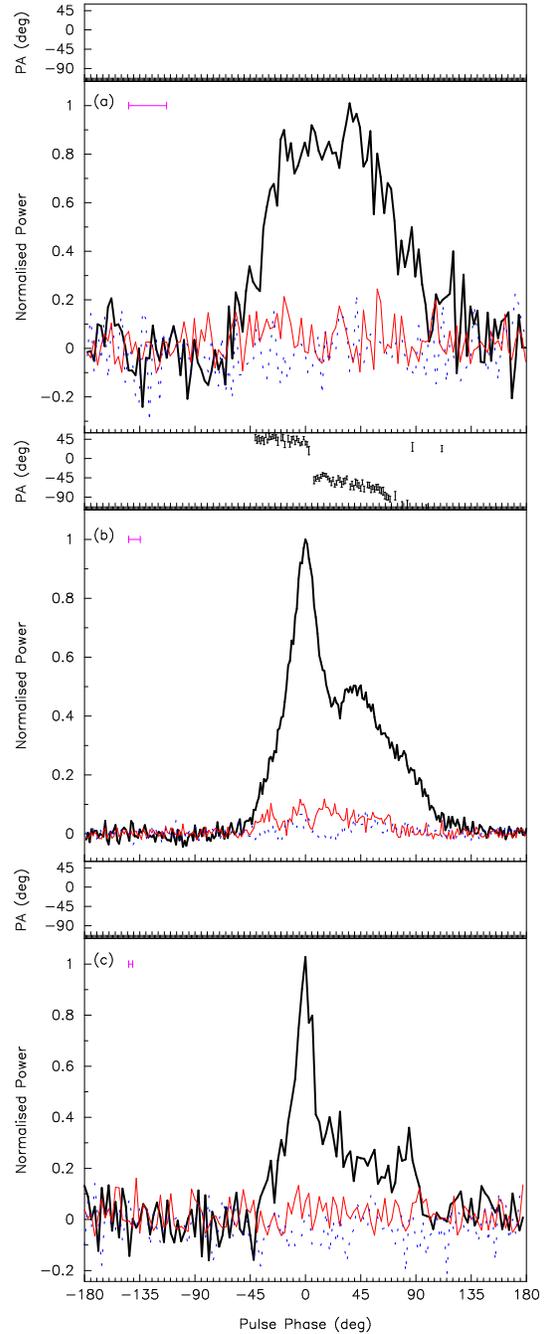}
\centering
\caption{
\label{J1708-3506}
Polarisation profiles of PSR J1708--3506 at (a) 732, (b) 1369 and (c) 3100~MHz.
}
\end{figure}

\polpsrsec{PSR J1708--3506} (Figure \ref{J1708-3506}).
The profile of PSR J1708--3506 appears to be formed of two main components, with the trailing component much broader than the leading.
The broad component appears to have a steeper spectral index, dominating the profile at 732~MHz and appearing only as a small shoulder at 3100~MHz, causing a relatively large profile evolution with frequency, as compared to most MSPs.
The level of polarisation is less than $10\%$ throughout, with the intensity of linear and circular following each other.
An orthogonal mode jump is apparent $\sim 5^\circ$ after the peak of the profile.
There appears to be a corresponding change in the handedness of circular polarisation associated with the jump, however this later reverses again without a corresponding change in PA.

\begin{figure}
\includegraphics[width=7cm]{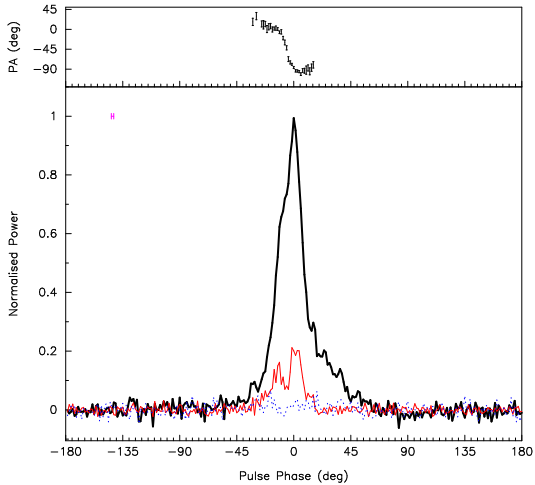}
\centering
\caption{
\label{J1719-1432}
Polarisation profile of PSR J1719--1438 at 1369~MHz.
}
\end{figure}

\polpsrsec{PSR J1719--1438} (Figure \ref{J1719-1432}).
The profile of PSR J1719--1438 is relatively narrow, and shows a high shoulder on the leading edge and a low shoulder on the trailing edge.
The linear polarisation fraction is $\sim 16\%$ without detectable circular polarisation.
The PA swing is steep with the steepest gradient leading the peak of the profile by $6^\circ$.

\begin{figure}
\includegraphics[width=7cm]{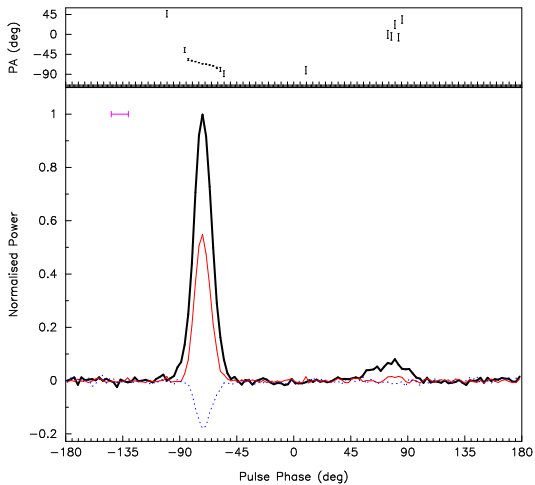}
\centering
\caption{
\label{J1731-1845}
Polarisation profile of PSR J1731--1845 at 1369~MHz.
}
\end{figure}

\polpsrsec{PSR J1731--1845} (Figure \ref{J1731-1845}).
The profile of PSR J1731--1845 is composed of two main components, separated by $155^\circ$.
The brightest component is polarised with a linear fraction of $50\%$ and circular fraction of $20\%$.
The second component is much less polarised, with only a hint of linear polarisation.

\begin{figure}
\includegraphics[width=7cm]{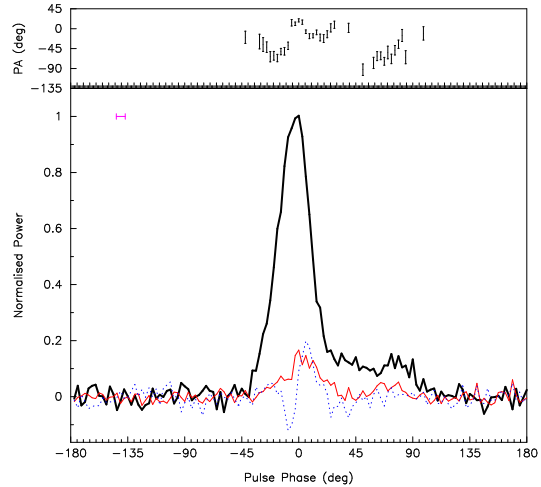}
\centering
\caption{
\label{J1801-3212}
Polarisation profile of PSR J1801--3212 at 1369~MHz.
}
\end{figure}

\polpsrsec{PSR J1801--3212} (Figure \ref{J1801-3212}).
The profile of PSR J1801--3212 consists of a bright leading component with a shoulder.
The emission is elliptically polarised with a linear and circular polarisation fraction of $\sim 15\%$ each.
The peak of the polarised intensity trails the peak of the total intensity by $\sim 18^\circ$.
The PA does not appear to follow the typical `S-shaped' swing predicted by the RVM.
The emission changes hand of circular polarisation at a phase corresponding to the peak of the total intensity profile.
Just prior to this phase, the PA of the linear polarisation jumps by $\sim 50^\circ$.
The limited S/N makes it hard to discern, but there may also be a true $90^\circ$ jump in the PA between the polarisation associated with the main peak and that of the shoulder, occurring at a phase of $\sim 40^\circ$.
There may also be a hint that handedness of the circular polarisation may also reverse around, or just preceding this phase.

\begin{figure}
\includegraphics[width=7cm]{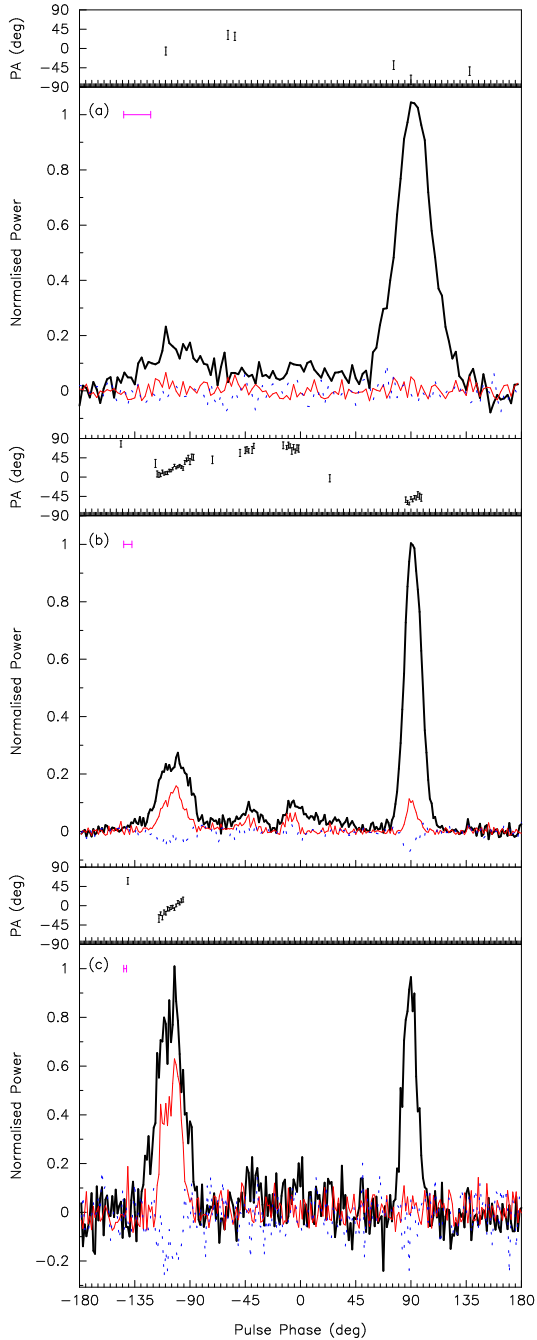}
\centering
\caption{
\label{J1811-2404}
Polarisation profiles of PSR J1811--2404 at (a) 732, (b) 1369 and (c) 3100~MHz.
}
\end{figure}

\polpsrsec{PSR J1811--2404} (Figure \ref{J1811-2404}).
At 1369~MHz we can discern at least four independent components with emission spread over $250^\circ$ of pulse phase.
The emission has a mean linear polarisation fraction of $16\%$ and a circular polarisation fraction of $12\%$.
Of the two brightest components, the leading component has a flatter spectral index and a higher polarisation fraction.
The trailing component dominates the profile at 732~MHz, and the two components are equal at 3100~MHz.
The profile appears to be almost completely unpolarised at 732~MHz, even though the leading component has a considerable polarisation fraction at 1369 and 3100~MHz.
The inner components are also polarised and form a bridge of emission between the two main components.

\section{Discussion of polarisation and speculation on geometry}
\label{sec_discuss}
We can use the RVM to model the observed polarisation PA $\Psi$ in terms of geometrical parameters $\alpha$ and $\beta$ by
\begin{equation}
\label{equ:rvm}
\tan (\Psi+\Psi_0) = \frac{ \sin \alpha \; \sin(\phi-\phi_0)}{
\sin\zeta\; \cos\alpha - \cos\zeta\sin\alpha
\cos(\phi-\phi_0)},
\end{equation}
where $\zeta = \alpha + \beta$, and $\Psi_0$ and $\phi_0$ are the fiducial phase and PA, nominally corresponding to the magnetic axis crossing the line of sight.
The relation between $\phi_0$ and the observed emission is however affected by aberration and retardation effects in the pulsar magnetosphere.
If the emission height, $r_{\rm em}$, is small compared to the light cylinder radius, $R_{\rm LC}$, then $\phi_0$ arrives later with respect to the corresponding total intensity emission, with a shift given by
\begin{equation}
\Delta\phi = 4r_{\rm em}/R_{\rm LC}
\label{eq_height}
\end{equation}
\cite{bcw91,ha01,dyk08}.

The magnetospheres of MSPs are very compact, a pulsar with a spin period of 5~ms has a light cylinder radius of only 240~km. Conventionally it is assumed that the radio emission occurs from a radius which is only a small fraction (few percent) of the light cylinder although in young pulsars higher emission heights are common \cite{jw06}. In MSPs, low emission heights cause a problem because an emission height of 10\% of the light cylinder yields a pulse width of only $\sim$60\degr\ unless the rotation and magnetic axes are close to alignment. Observationally it has been clear for over a decade that MSP widths are much larger than this on average \cite{xkj+98,ymv+11}. Indeed, the observed large widths preclude any emission height less than the light cylinder in the conventional picture, unless the majority of MSPs are aligned rotators. Conversely, for a number of MSPs where there has been an attempt to constrain the geometry, it has been found that the inferred opening angle is much smaller than that expected by extrapolation from the main population of radio pulsars \cite{kxl+98}.

In principle, the polarisation properties could help shed light on this conundrum, especially as large profile widths are a boon to RVM fitting. Unfortunately, the PA variations in MSPs can be extremely complex and are often far from that expected in the RVM. Furthermore, the lack of obvious inflexion point in the PA variations and the lack of identifiable core/cone components makes it difficult to estimate emission heights via the aberration/retardation methods of \citeN{bcw91}. We note in passing however that MSPs appear to be no better or worse than normal pulsars in this regard \cite{ew01}. Indeed the geometry of the wide profile pulsars B0950+08, B1822$-$09 and B1929+10 to take but three examples remains unclear \cite{ew01}. In spite of these inherent difficulties we have attempted RVM fits to several of the MSPs in our sample in an attempt to get a handle on geometry and emission locations.

\subsection{PSR~J1017--7156}
\label{1017_discuss}
Neither the drastic change of polarisation with frequency, nor the `U' shaped PAs observed in this pulsar can be simply explained within the framework of the RVM.
We also observe a change in the handedness of circular polarisation, and a $\sim 90^\circ$ change in the absolute PA of the linear polarisation between the profile at 732~MHz and 1369~MHz.
We have considered the possibility that this effect is caused by an error in calibration, however multiple observations of this pulsar and the lack of similar effects in other pulsars largely rules this out.
This is, therefore, a strong suggestion that the profile is composed of two competing elliptically polarised orthogonal modes, with different spectral indices \cite{sse+06}.
This superposition of modes can also explain the rapid sweeps in PA observed at the trailing edge of the pulse at 732 and 1369~MHz.
The strong depolarisation of the profile at the trailing edge suggests that there is a change in the dominant polarisation mode at this phase, and the PA should jump by 90$^\circ$.
Broadening of the pulse in the interstellar medium can smooth these discrete jumps somewhat, causing the appearance of a rapid swing in PA \cite{nkk+09,kar09}.

\subsection{PSR~J1125--5825} 
\begin{figure}
\includegraphics[width=7cm]{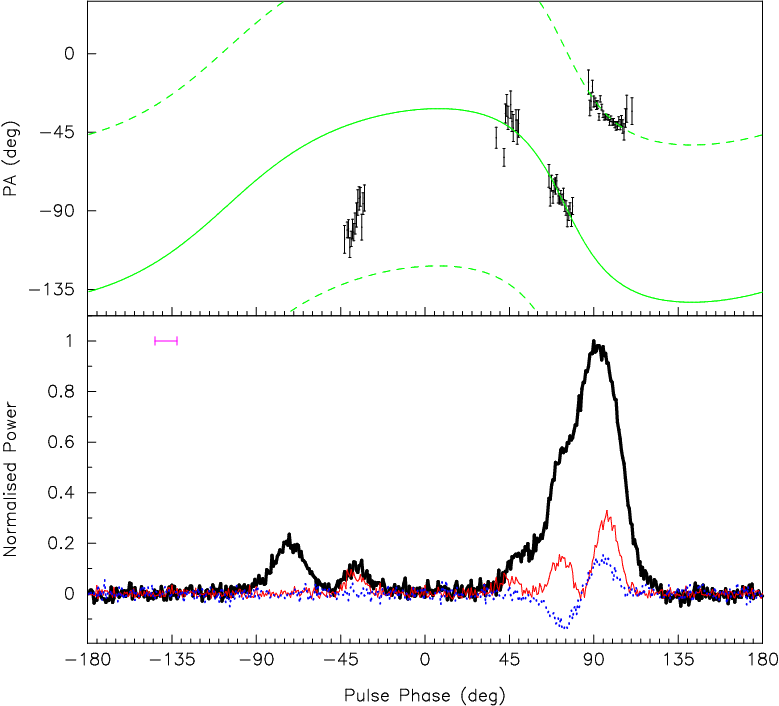}
\centering
\caption{
\label{J1125_rvm}
Polarisation profile of PSR J1125--5825 at 1369~MHz, showing the observed PAs overplayed with the model obtained from fitting the RVM.
The dashed line is separated by 90$^\circ$ from the model PA to indicate orthogonal modes.
}
\end{figure}

The `main pulse' of this pulsar extends over 100\degr\ of longitude and has a well behaved swing of polarisation once the orthogonal jump near the centre of the profile is taken into account. The presence of a further polarised component some 130\degr\ away from the main pulse is an added complication; it is not possible to fit the RVM to all the data. However, it is possible to obtain a good RVM solution using only the main pulse data. In this case, $\alpha=128\degr$, $\beta=-21\degr$ with the inflexion point coincident with the negative peak of the circular polarisation. The RVM correctly predicts the swing of the polarisation at the distant component but the values of PA are $\sim$30\degr\ offset from the observed values.
The data and model are shown in Figure \ref{J1125_rvm}.

We note also the similarity between this profile and that of the MSP ~J1012+5307 \cite{xkj+98, stc99} and also with the normal pulsar B1055--52 \cite{ww09}. In the former case the authors were unsure whether the pulsar is an aligned or an orthogonal rotator whereas the latter appears to be clearly orthogonal. For PSR~J1125--5825 the ambiguity remains due to the problematical nature of the RVM fit. If the pulsar is an aligned rotator this could explain the large width and the complex RVM can arise if different components emit at different heights. On the other hand, if the pulsar is an orthogonal rotator the likely implication is that the main and interpulse emission arise from significantly different heights or indeed from the outer magnetosphere, resulting in a non-standard RVM.
With more time, and therefore greater S/N, the gamma-ray profile of PSR~J1125--5825 may be able to provide further constraints on the geometry, and clues to the location of the radio emission regions.

\subsection{PSR~1502--6748}
This pulsar appears to be a very wide triple profile with the central
peak dominating, and two weak outriders located some 80\degr\ away.
The PA swing is smooth, the RVM fit is not very constraining on $\alpha$
and $\beta$ but does locate the inflexion point some 40\degr\ distant
from the main peak.  This yields a height of 220~km, about 20\% of
the light cylinder radius, using Equation \ref{eq_height}. Although this seems plausible it is difficult
to reconcile this relatively low height with the large pulse width
unless $\alpha$ is small.

\subsection{PSR~J1708--3506}
The emission from this pulsar seems to conform to standard behaviour.
It is a blended triple, with the central component having a steep spectral index and located equidistant from the flatter spectrum conal components.
The PA swing is relatively smooth apart from an orthogonal jump. The RVM
fit is reasonably well constrained with $\alpha$ between 60\degr\ and
90\degr\ and $\beta$ of 20\degr. The inflexion point occurs some 40\degr\
after the location of the central component. If we interpret the offset
as due to aberration, this yields a height of $\sim$40~km, which, assuming an emission region filled to the edge of the open field line region, would imply a profile width of $\sim$80\degr.
The observed profile is however much broader than this, placing doubt on the RVM-derived geometry, or otherwise suggesting a more complex model of the emission region is required.
The large frequency-dependant evolution of the profile may suggest that the profile is composed of components that are originating in different parts of the magnetosphere and therefore confusing our picture.

\subsection{PSR~J1719--1438}
This pulsar exhibits an almost classical S-shaped swing of linear polarisation across the pulse.
The profile is again extremely wide, and this width coupled with the PA swing implies a small value of both $\alpha$ and $\beta$.
The inflexion point of the traverse is located some 6\degr\ prior to the pulse peak.
The geometry of this star is thus rather uncertain and cannot shed much light as to the origins and evolution of its planetary mass companion \cite{bbb+11b}.

\subsection{PSR~J1811--2404}
\begin{figure}
\includegraphics[width=7cm]{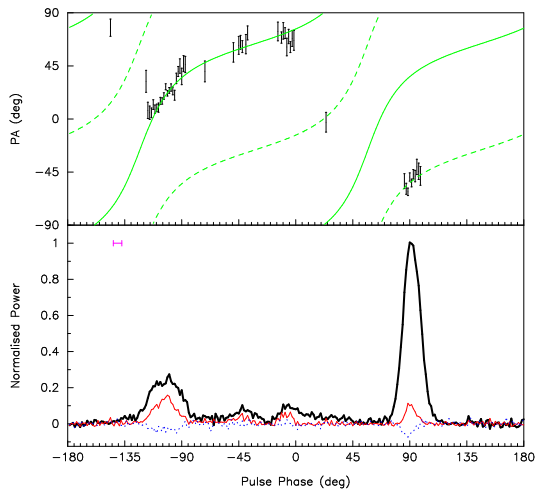}
\centering
\caption{
\label{J1811_rvm}
Polarisation profile of PSR J1811--2404 at 1369~MHz, showing the observed PAs overlaid with the model obtained from fitting the RVM.
The dashed line is separated by 90$^\circ$ from the model PA to indicate orthogonal modes.
}
\end{figure}

Polarised emission extends over nearly 300$\degr$ of longitude in four
distinct components. By inserting an orthogonal mode jump between the
dominant component and the weaker components we can obtain a remarkably
good and strongly constrained RVM fit with $\alpha$=89.7, $\beta$=21
and the location of the inflexion point 145$\degr$ after the main pulse.
The data and model are shown in Figure \ref{J1811_rvm}.
The pulsar therefore seems to be an orthogonal rotator where we see
emission from both poles. The question remains as to which components belong to which pole, and which (if any) are centrally located.
There are two possibilities.

The first option is that the core is located at phase $90\degr$ and $-90\degr$.
The main pulse would then consist only of the core component with no
conal outriders. The interpulse consists of the core plus two trailing
conal components. The spectral index evolution does not fit this picture
very well, and the extreme width of the interpulse is also difficult to
reconcile.
An alternative form of this picture would assign the two central components to a different emission height, or from an emission region located in the outer magnetosphere.

In the second possibility, the core is located at phase
$135\degr$ and $-45\degr$. In this case, the main pulse consists only of
a leading edge cone whereas the interpulse is a mostly symmetric triple
structure. Under this picture the offset between the inflexion point and
the core location yields a height of 54~km (c.f. the light cylinder radius
of 124~km) and this height is consistent with a pulse width of some
120$\degr$, about that of the interpulse emission.

\subsection{Summary}
There now exists a substantial body of high quality polarisation measurements of MSPs built up over the last decade. Key points to re-iterate
from these observations are (a) many MSPs have large widths especially
when observed with high dynamic range \cite{ymv+11}, (b) as in normal pulsars a
wide range of linear and circular polarisation fractions are observed
(c) the polarisation profiles can be highly complex and the PA swing often does not obey RVM, (d) a non-negligible fraction are amenable to RVM fitting.

It seems difficult to reconcile the observations with a low emission height
assuming an isotropic distribution of $\cos(\alpha)$.
However, it may be that $\alpha$ is biased towards small values in MSPs,
linked to their formation process. This has to be investigated further but the
presence of bridges of emission between the various components might support this idea.  However, it is also not a given from the data that the entire phenomenological view of the radio emission has to be overturned in MSPs by invoking e.g. caustic emission or extreme magnetosphere effects \cite{dwd10}.
Consider a range of allowable emission heights up to $\sim$50\% of the light cylinder radius. For isotropic values of $\cos(\alpha)$ one would
then expect minimum pulse widths of $\sim$50\degr\ and high fraction
(30\%) of interpulses. This is more or less consistent with the data where
it could be argued that at least 3 of our sample and 7 of the \citeN{ymv+11}
sample show evidence of interpulse emission. In these interpulse pulsars there is no {\it a priori} reason why either the flux density of the emission
from the two poles should be similar (e.g the extreme case of 
PSR J1603$-$7202) nor that the emission heights should be similar which would
lead to highly complex PA traverses. The bridge emission seen in
e.g. PSRs 1939+2134 and 1045$-$4509 is not evidently part of the interpulse
picture but the wide beams mean that naturally there will be blending
between the main and interpulse profiles, so we do not see this as a major issue.  For the pulsars which clearly
do not show interpulse emission, some have relatively narrow profiles
(PSRs 1017$-$7156, 1022+1001, 1446$-$4701, 1600$-$3053, 1713+0747) whereas some are wide (PSRs 1543$-$5149, 1708$-$3506, 1732$-$5049)
perhaps indicating lower values of $\alpha$.
In summary, we believe that emission heights in MSPs are high, up to 50\% of the light cylinder.
This leads to a substantial fraction of MSPs with interpulses, further complicating the identification of components in the profile and the PA swing.

Although some MSPs have polarisation consistent with the RVM, it is still not clear that the derived geometry can be used to determine a consistent model for the emission location, and it is possible that the accretion process itself has modified the magnetosphere to a point where a simple dipolar model is no longer applicable \cite{xkj+98}.
Overall, however, we find that the polarised emission of MSPs is not more complex than that observed in many slow pulsars.
There has been success in untangling these complex PA variations in slow pulsars through studies of polarised intensity on a pulse-by-pulse basis (e.g. \citealp{br80,gl95,kkj+02}).
Future high-sensitivity observations may well be able to repeat this success in MSPs, giving us greater confidence in geometric interpretations of the RVM and a fuller picture MSP emission.
High significance gamma-ray profiles of MSPs could also provide additional constraints on geometry, and test theories of radio emission from the outer magnetosphere.

\section{Conclusion}

The High Time Resolution Universe survey for pulsars and fast transients has discovered 12 MSPs to date, six of which are announced in this work.
These six MSPs (indeed, all 12) are in binaries.
Two, PSRs J1017-7156 and J1543-5149, with spin periods of $\sim 2$~ms, binary periods of about 1 week and companions with $m_{\rm c,min}\sim 0.2$~M$_\odot$ are easily classified as LMBPs, the most common category of MSPs.
PSR J1337--6423, with its larger companion mass ($m_{\rm c,min}=0.8$~M$_\odot$) and longer spin period (9.4~ms) better fits into the class of IMBPs, likely with a heavy WD companion.
In contrast, PSR J1446--4701 has a much lower mass companion ($m_{\rm c,min}=0.02$), falling easily into the distinct category of VLMBPs, typified by the `black widow' pulsars.
The high rotational energy loss rate, $\dot E \sim 4\times 10^{34}$, and inferred distance of only 1.5~kpc suggests that this is a good candidate for detection as a gamma-ray pulsar.
Indeed, we find a tantalising indication of gamma-ray pulsations by using the radio ephemeris to fold gamma-ray photons detected by the {\it Fermi} LAT.

The final two pulsars, PSRs J1502--6752 and J1622--6617, both have spin periods of $\sim 25$~ms, typically considered intermediate spin periods for recycled pulsars.
PSR J1622--6617 can be classified as an intermediate period LMBP, similar to other known systems such as PSR J1841+0130.
There are however no similar systems to PSR J1502--6752 in the literature.
The mass function implies $m_{\rm c,min}=0.02$, typical of the VLMBPs, however the formation of these systems is thought to spin the pulsar up to very short periods.
Whilst it is possible that the low mass function is an inclination effect, it is certainly striking that there are no MSP systems with $0.026 < m_{\rm c,min} < 0.09$, with a large number of MSPs having $m_{\rm c,min}\sim 0.1$ or $m_{\rm c,min}\sim0.02$.
Therefore we conclude that the companion of PSR J1502--6752 is likely to have formed through the same channel as the companions for the VLMBPs, and so that formation mechanism does not require, or does not guarantee, a short spin period MSP.

We have also undertaken polarimetric observations of all 12 of the MSPs discovered by the HTRU survey to date.
The calibrated profiles show the wide variety of profiles typical of MSPs, with pulse widths ranging from $24^\circ$ to $280^\circ$ (measured at $10\%$ of peak flux).
The profiles generally consist of multiple components, often showing broad shoulders running off either the leading or trailing edge of a narrow central feature.
PSRs J1125--5825, J1731--1855 and J1811--2404 show the dominant components separated by $\sim 180^\circ$, and may either be classified as interpulse pulsars or extremely wide cones.
For a number of the MSPs, we find that the observed swing in PA can be fit by the RVM, however there are clearly cases such as PSR J1017--7156 where effects such as smoothed orthogonal mode jumps make this impossible.
We believe that emission heights in MSPs are a substantial fraction of the light cylinder, leading to a large fraction of MSPs showing emission from both poles.

\section{Acknowledgements}
The Parkes Observatory is part of the Australia Telescope which is funded by the Commonwealth of Australia for operation as a National Facility managed by CSIRO.

The \textit{Fermi} LAT Collaboration acknowledges generous ongoing support
from a number of agencies and institutes that have supported both the
development and the operation of the LAT as well as scientific data analysis.
These include the National Aeronautics and Space Administration and the
Department of Energy in the United States, the Commissariat \`a l'Energie Atomique
and the Centre National de la Recherche Scientifique / Institut National de Physique
Nucl\'eaire et de Physique des Particules in France, the Agenzia Spaziale Italiana
and the Istituto Nazionale di Fisica Nucleare in Italy, the Ministry of Education,
Culture, Sports, Science and Technology (MEXT), High Energy Accelerator Research
Organization (KEK) and Japan Aerospace Exploration Agency (JAXA) in Japan, and
the K.~A.~Wallenberg Foundation, the Swedish Research Council and the
Swedish National Space Board in Sweden.  
Additional support for science analysis during the operations phase is gratefully
acknowledged from the Istituto Nazionale di Astrofisica in Italy and the Centre National d'\'Etudes Spatiales in France.

The authors would like to thank P. Ray for help with the gamma-ray analysis section.

\bibliographystyle{mnras}
\bibliography{journals,myrefs,modrefs,psrrefs,crossrefs}

\end{document}